\documentclass[pdflatex,sn-mathphys-num]{sn-jnl}

\usepackage[utf8]{inputenc}
\usepackage{graphicx}
\usepackage{multirow}
\usepackage{amsmath,amssymb,amsfonts}
\usepackage{amsthm}
\usepackage{mathrsfs}
\usepackage[title]{appendix}
\usepackage{xcolor}
\usepackage{textcomp}
\usepackage{booktabs}
\usepackage{algorithm}
\usepackage{algorithmicx}
\usepackage{algpseudocode}
\usepackage{listings}
\usepackage[numbers]{natbib}
\usepackage{textgreek}
\usepackage[mathscr]{euscript}
\let\mathscr\EuScript
\theoremstyle{thmstyleone}

\usepackage{lineno}
\theoremstyle{thmstyletwo}

\theoremstyle{thmstylethree}

\begin{document}

\title[Article Title]{Nonlinear synchronization through vector subharmonic entrainment}

\author*[1]{\fnm{Dmitrii} \sur{Stoliarov}}\email{d.stoliarov@aston.ac.uk}
\equalcont{These authors contributed equally to this work.}
\author[1]{\fnm{Sergey} \sur{Sergeyev}}\email{s.sergeyev@aston.ac.uk}
\equalcont{These authors contributed equally to this work.}
\author[1]{\fnm{Hani} \sur{Kbashi}}\email{h.kbashi@aston.ac.uk}
\author[2]{\fnm{Fan} \sur{Wu}}\email{fanwu@shu.edu.cn}
\author[2]{\fnm{Qianqian} \sur{Huang}}\email{huangqq@shu.edu.cn}
\author*[2]{\fnm{Chengbo} \sur{Mou}}\email{mouc1@shu.edu.cn}
\equalcont{These authors contributed equally to this work.}

\affil*[1]{\orgdiv{Aston Institute of Photonics Technologies}, \orgname{Aston University}, \orgaddress{\street{}, \city{Birmingham}, \postcode{B4 7ET}, \country{UK}}}
\affil[2]{\orgdiv{Key Laboratory of Specialty Fiber Optics and Optical Access Networks, Joint International Research Laboratory of Specialty Fiber Optics and Advanced Communication }, \orgname{Shanghai University}, \orgaddress{\street{}, \city{Shanghai}, \postcode{200444}, \country{China}}}

\abstract{
Synchronization is ubiquitous across a wide range of fields. Subharmonic entrainment (SHE) is a nonlinear synchronization phenomenon that results in a locking oscillator at a frequency of an external periodic forcing signal with a fraction of the oscillator frequency. Beyond the fundamentals of nonlinear dynamics, SHE has a range of practical applications—from stabilizing ultrafast laser pulses to optimizing control in various engineering and natural systems. However, the vectorial nature of SHE remains elusive. Here, we present the results of a theoretical and experimental study of a vector type of subharmonic entrainment (VSHE) using a passively mode-locked fiber laser as a testbed. We unveil the mechanism of vectorial SHE, in which weak external signals can entrain internal laser dynamics through vectorial coupling. Vectorial SHE presents in the form of synchronization between the subharmonic of mode-locking-driven oscillations and continuous wave (CW) signal through an evolving state of polarization. This CW signal, driven by the internal dynamics of the injected signal, causes VSHE with the frequencies’ ratios of multiples of ten, resulting in a partially mode-locking regime operation. Our findings offer new control techniques over mode-locking and additional dimension such as polarization states.}

\maketitle

\section{Introduction}\label{sec1}
Synchronization phenomena, such as those described by coupled oscillator models—where multiple interacting systems spontaneously coordinate their dynamics—are fundamental in various natural and engineered networks\cite{bialek2012,papo2014,schintler2005,batool2017,okeeffe2017,mancini2010,minati2019,arenas2008,pikovsky2001}. Typically, the research agenda focuses on revealing mechanisms of how the interactions between individual oscillators driven by localized external perturbations caused by the injected signal result in synchronization (injection locking), e.g., the capability of coupled oscillators to synchronize at a common frequency \cite{bialek2012,papo2014,schintler2005,batool2017,okeeffe2017,mancini2010,minati2019,arenas2008,pikovsky2001}. For example, in the field of photonics, injection locking of lasers was initially motivated by the ability to stabilize high-power continuous wave lasers by using low-power master lasers \cite{buczek1973,lang1982}. Subsequent studies have investigated various injection locking techniques for stabilizing pulsed lasers such as mode-locked lasers' pulse trains \cite{gat2013,yu2022}, multi-wavelength solitons \cite{mao2021}, vector soliton rain dynamics control \cite{sergeyev2022}, bound state solitons \cite{chang2022,zou2022}, and both tunable \cite{kielpinski2012,ribenek2021a,ribenek2021b} and low-noise harmonic mode-locked pulses \cite{korobko2022a,korobko2023}. In particular, various synchronization scenarios driven by internal dynamics have been explored. These include frequency locking with free-running phase \cite{thevenin2011}, subharmonic entrainment of pulsating single solitons in an ultrafast laser leading to the breathing dynamics \cite{xian2020,wu2023}, synchronization of polarization modes with zero-log, phase difference entrainment and desynchronization \cite{sergeyev2023,kbashi2019,kbashi2018,sergeyev2021,sergeyev2014,sergeyev2014b,wang2023,wu2023b,wang2023b}. 

SHE is a form of synchronization in which phase locking occurs between coupled oscillators or between an oscillator and an injected signal \cite{pikovsky2001,wu2025unveiling}. Theoretically, SHE arises when an oscillator with a  frequency ${\omega}$  synchronizes to a lower frequency ${\omega_0}$  typically satisfying a rational ratio ${\omega_0}/{\omega} < 1$. These synchronization regimes, known as Arnold tongues, allow phase locking even under small detuning, provided the coupling strength is sufficient. In particular, the synchronization region for SHE with ${\omega_0}/{\omega} = 0.5$  can be relatively broad \cite{pikovsky2001}.

The spring-coupled simple pendulum system Fig.~\ref{fig:1} a provides a classical analogue to the dynamics of orthogonally polarized modes in mode-locked lasers, where each pendulum represents an oscillator associated with one polarization state. The spring introduces coherent coupling, enabling energy exchange and phase synchronization—similar to the interaction between polarization components in vector mode-locked lasers. In the strong coupling regime, the pendulums exhibit hybridized normal modes with in-phase and anti-phase oscillations, leading to frequency splitting or level repulsion. This behavior parallels the formation of stable vector solitons, where the phase difference between polarization states remains locked. Conversely, when the coupling is weakened or detuning exceeds the locking bandwidth, the pendulums desynchronize, analogous to polarization mode desynchronization in lasers, resulting in dissipative soliton breathing \cite{wu2023}. The evolution of phase difference in both systems is well captured, highlighting the transition between synchronization, entrainment, and chaotic dynamics. Thus, the spring-coupled pendulum serves as an intuitive mechanical model for understanding phase dynamics and coupling-induced transitions in vector mode-locked laser systems.

In this work, we report for the first time the experimental and theoretical demonstration of vector subharmonic entrainment (VSHE) in an ultrafast laser. We experimentally observe locking between a low-frequency injected polarization-modulated signal and internal polarization oscillations in an ultrafast laser. This phenomenon emerges as a double-timescale pulsation regime: a train of ultrashort pulses modulated by a slowly varying envelope, both synchronized to an externally injected CW signal with deterministic polarization modulation. Our results reveal how internal nonlinear polarization dynamics can be harnessed via external low-frequency injection. The obtained stable partially mode-locking regime, also known as Q-switched mode-locking (QSML) with synchronized polarization evolution featuring a two-dimensional high-order synchronization of both temporal and polarization domains, gaining novel insights into vectorial synchronization in nonlinear optical systems.  
\section{Results}\label{sec2}
\textbf{Experimental demonstration }

\begin{figure}[h]
\centering
\includegraphics[width=1.0\textwidth]{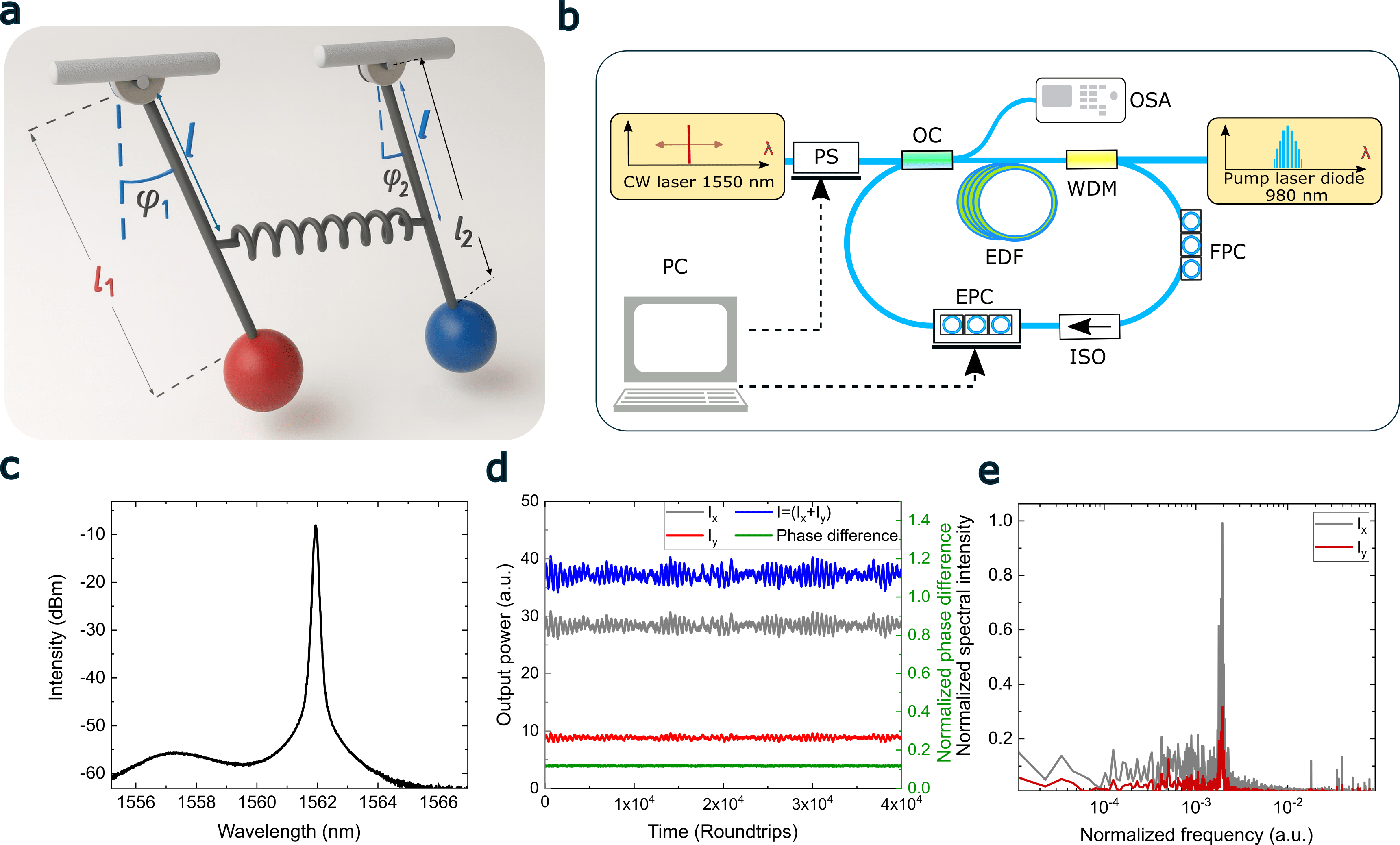}
\caption{
Principle and schematic of subharmonic entrainment induced nonlinear synchronization: (a)  Simple spring-coupled oscillator model: two pendulums of different mass and unequal lengths $l_1$ and $l_2$ are coherently coupled via a spring of finite stiffness. The angular displacements $\varphi_1$ and $\varphi_2$ describe the phase evolution of each oscillator, which possess distinct natural frequencies.
(b) Schematic setup of the NPR mode-locked fiber laser. EDF: erbium-doped fiber;  980~nm pump laser diode;1550 nm continuous wave laser: CW laser 1550 nm; FPC: manual fiber polarization controller; EPC: electronically driven polarization controller; ISO  polarization-sensitive optical isolator; WDM: wavelength-division multiplexer; PS: polarization scrambler;  OC: 70/30 optical coupler, OSA: optical spectrum analyzer
(c) Optical spectrum of the CW regime of the NPR mode-locked Er-doped fiber laser.
(d) Polarization dynamics of the NPR mode-locked Er-doped fiber laser in terms of oscillations of the output powers of the individual polarization components $I_x$ (blue) and $I_y$ (green), total power $I = I_x + I_y$ (red), and the phase difference $\Delta\varphi$ between orthogonal $x$- and $y$-SOPs normalized to $\pi$ (black).
(e) Fast Fourier transform spectrum of the $x$- and $y$-SOPs’ power dynamics with frequencies normalized to the fundamental frequency $f_0 = 16.67~\text{MHz}$.
Parameters: pump power $I_p = 72~\text{mW}$, averaging time for the polarimeter trace $T_{\text{pol}} = 320~\text{ns}$.}
\label{fig:1}
\end{figure}

To investigate VSHE, we built a nonlinear polarization rotation (NPR)-based mode-locked fiber ring laser operating under the anomalous dispersion regime. The developed test bed is well suited for studying internal nonlinear polarization dynamics within a fiber laser cavity. In this setup, a QSML regime is observed, which could be driven by an internal and external  CW signal. This regime exhibits two-dimensional high-order partial synchronization across both temporal and polarization domains. Such behavior provides a robust experimental platform for investigating vectorial synchronization in nonlinear optical systems. A schematic representation of the mode-locked fiber laser configuration  is shown in Fig.~\ref{fig:1} b, the detailed description of the setup is provided in the Methods section. 

We seeded an optical power of $15\,\text{dBm}$ at $1.67\,\text{kHz}$ through a vacant input port of a 30/70 output optical coupler. A fast polarimeter PM1000-XL-FA-N20 D (Novoptel) with a sampling frequency of 100 MS/s to observe the evolution of the Stokes parameters $S_1, S_2$, and $S_3$, the power for the orthogonal x- and y- polarization components (Ix, Iy), the total power $I$, and the phase difference  $\Delta\varphi$ was recalculated from the measured Stokes parameters as follows:  

\begin{equation}
\begin{aligned}
I &= I_x + I_y, \quad S_1 = I_x - I_y, \\
S_2 &= 2\sqrt{I_x I_y} \cos(\Delta\varphi), \quad
S_3 = 2\sqrt{I_x I_y} \sin(\Delta\varphi).
\end{aligned}
\label{eq:stokes}
\end{equation}

The polarization resolved results of our experimental study are shown in Figs.~\ref{fig:1} c-e. During the experiments, the pump power was maintained at a constant level of $I_p = 72\,\text{mW}$, with only the polarization controllers being adjusted.
In the absence of the injected signal, we tuned the electronically driven polarization controller (EPC) to establish the CW regime, as shown in Fig.~\ref{fig:1} c. As illustrated, the optical spectrum exhibits a localized, narrowband spike-like structure centred around 1562 nm. The polarization resolved temporal behaviours of pulses, shown in Fig.~\ref{fig:1} d, appear as low-amplitude oscillations in the output powers of the orthogonal states of linear polarization, $I_x$ and $I_y$, the total power $I = I_x + I_y$, and the phase difference $\Delta\varphi$. A fast Fourier transform reveals that the oscillation frequency is $f_{\text{CW}} = 2 \times 10^{-3} \cdot f_0$ for the $x$- and $y$-SOPs, as shown in Fig.~\ref{fig:1} e. Given that the phase difference in Fig. 1 d is constant, we have a typical case of a phase-locked regime [9].

\begin{figure}[h]
\centering
\includegraphics[width=0.9\textwidth]{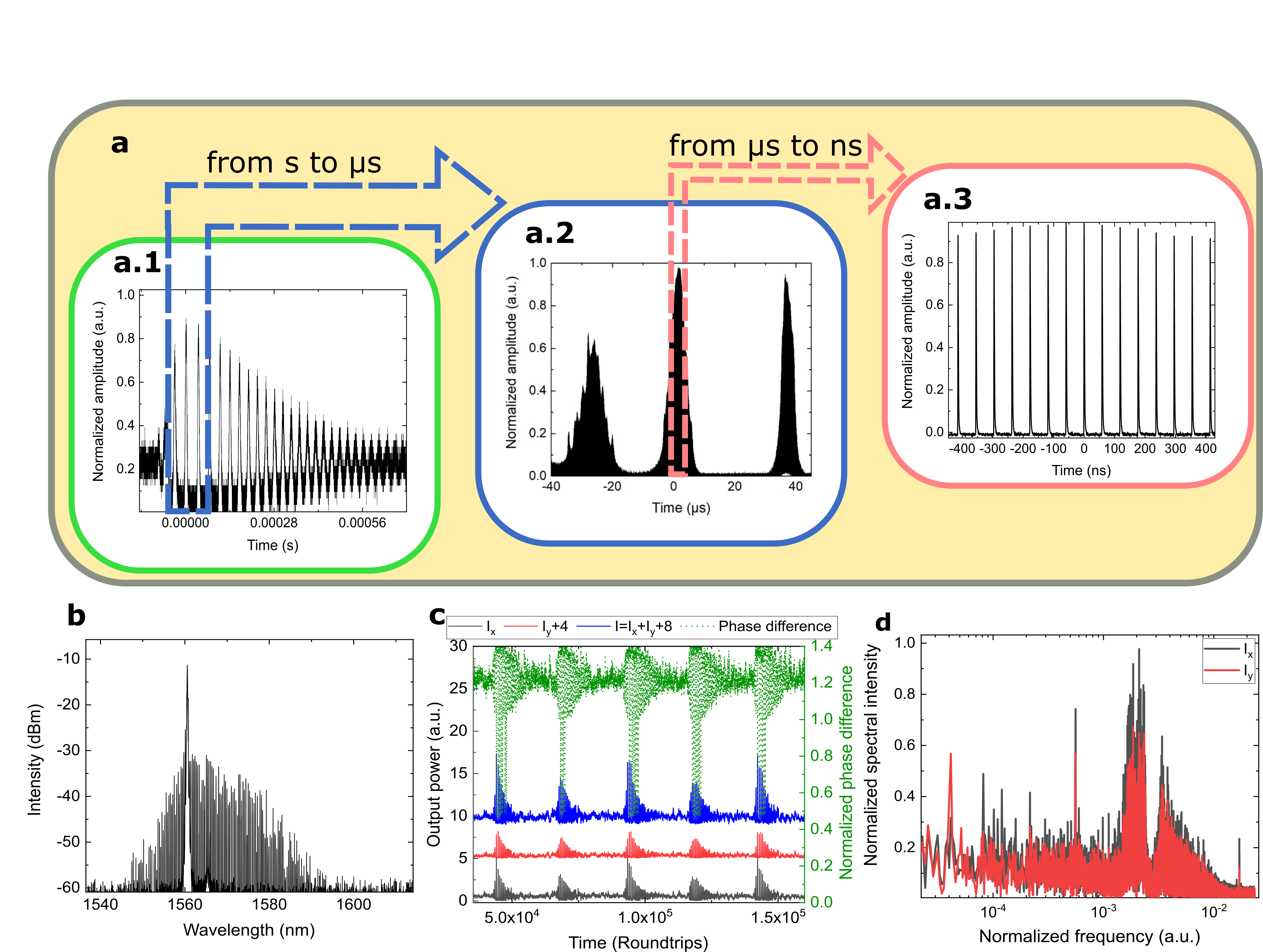}  
\caption{
\textbf{Partially mode-locking regime of ultrafast laser.} (a) Oscillogram at different timescales: a.1 second scale, a.2 microsecond scale, a.3 nanosecond scale;(b) Optical spectrum;(c) Polarization dynamics;(d) Fast Fourier transform spectrum (see notations in Fig.~\ref{fig:1}).\\The averaging time for the polarimeter trace is $T_{\text{pol}} = 1.28~\mu\text{s}$.}
\label{fig:2}
\end{figure}

By further adjusting the polarization controller, we were able to induce QSML dynamics (Figs.~\ref{fig:2} a.1 - a.3 ). As seen in Fig.~\ref{fig:2} b and Fig.~\ref{fig:2} d, the optical spectrum indicates the presence of a CW component with the peak wavelength at 1562 nm and frequency 
$f_{\text{CW}} = 2 \times 10^{-3} \cdot f_0$, similar to the previous case shown in Figs.~\ref{fig:1} c and \ref{fig:1} e.

The structure of a single fringe in the time domain and the partial mode-locking regime is shown in Fig.~\ref{fig:2}(a) and its inset. Slow QSML polarization dynamics are illustrated in Fig.~\ref{fig:2} c, where the dynamics exhibit double-scale oscillations in the output powers of the orthogonal polarization components: $I_x$ (black), $I_y$ (red), and total power $I = I_x + I_y$ (blue), along with the normalized phase difference $\Delta\varphi$ showing phase slips of approximately $\pi$ radians (green).

A fast Fourier transform reveals distinct spectral components in the power dynamics of the $x$- and $y$-SOPs: low-frequency oscillations at 
$f_{\text{LQ1}} = 4 \times 10^{-5} \cdot f_0$ and $f_{\text{LQ2}} = 8 \times 10^{-5} \cdot f_0$, and a high-frequency component at 
$f_{\text{HQ}} = 2 \times 10^{-3} \cdot f_0$ with multiple sidebands. The presence of sidebands in the high-frequency oscillations may contribute to overlap with the low-frequency components, thereby enabling partial synchronization through the SHE mechanism.

\begin{figure}[h]
\centering
\includegraphics[width=0.9\textwidth]{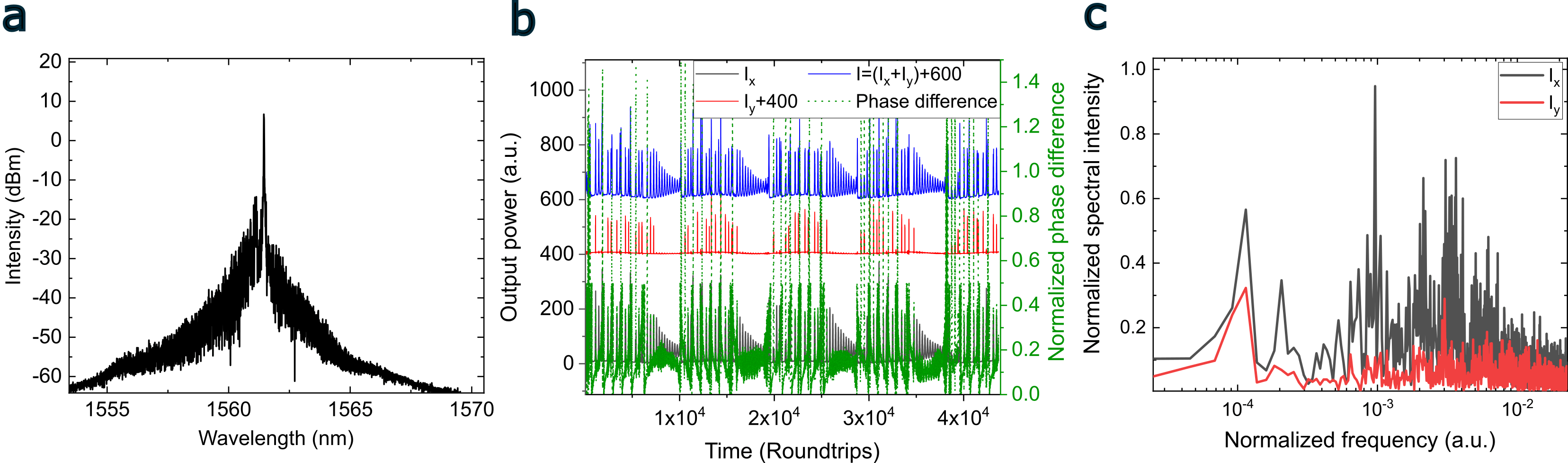}  
\caption{
\textbf{Vectorial subharmonic entrainment synchronized regime of ultrafast laser.} (a) Measured optical spectrum; (b) Polarization dynamics;
(c) Fast Fourier transform spectrum (see notations in Fig. 1). The averaging time for the polarimeter trace $T_{pol}=320\ ns$. 
}
\label{fig:3}
\end{figure}

To further investigate the VHSE dynamics driven by a low-frequency CW component in the cavity, an external CW laser with a deterministic state of polarization was used to inject a modulated low-frequency signal at 1562 nm. The central wavelength of the injected signal was detuned to match the central wavelength of self-started internal CW generation within the laser cavity, consistent with the configuration shown in Fig.~\ref{fig:1} c and Fig.~\ref{fig:2} b.
The results, shown in Fig.~\ref{fig:3} a–c, align closely with those in Fig.~\ref{fig:2} in terms of optical spectrum, polarization-resolved temporal dynamics, and the Fourier spectral components. Similarly to the dynamics in Fig.~\ref{fig:2} d, the slow QSML dynamics again reveal two characteristic frequency scales:
$f_{\text{LQ1}} = 10^{-4} \cdot f_0$, $f_{\text{HQ1}} = 10^{-3} \cdot f_0$, and $f_{\text{HQ2}} = 2 \times 10^{-3} \cdot f_0$.

\textbf{Theoretical Analysis}
To theoretically analyze the results, we implemented a vector model of an NPR-based Er-doped mode-locked fiber laser with an injected signal (see method section).
An approach which was previously used for theoretical characterization of NPR-based mode-locking is based on the introduction of function q which is related to amplitudes of the polarization components $u$ and $v$ as follows \cite{ding2009}:
   \begin{equation}
\begin{aligned}
u &= q \cdot \cos(\alpha_p), \\
v &= q \cdot \sin(\alpha_p).
\end{aligned}
\end{equation}

Here, $\alpha_p$ is the angle between the polarization plane of the polarizer and the slow axis~\cite{ding2009}. The complex amplitudes of the orthogonally polarized states of polarization (SOPs), $u$ and $v$, can also be expressed as follows:
\begin{equation}
\begin{aligned}
u &=|u| \exp(i \cdot \varphi_x),\\
\quad v &= |v| \exp(i \cdot \varphi_y).
\end{aligned}
\label{eq:3}
\end{equation}

As follows from Eqs.~(2) and (3), the phase difference is $\Delta\varphi = \varphi_y - \varphi_x \equiv 0$, and the polarization angle satisfies $\tan(\alpha_p) = \ {|v|}/{|u|}$.
However, as follows from Figs. 2 c and 3 b, the phase difference $\Delta\varphi$ is oscillating, which makes Eq. (2) not acceptable for the modelling of the polarization dynamics observed in our experiments. 

To overcome the drawbacks of previously used model \cite{ding2009} and so reveal mechanism of the mode-locking dynamics driven by the SHE between low-frequency oscillations, we updated a vector model of an Er-doped mode-locked fiber laser recently developed by Sergeyev and co-workers (details are found in \cite{sergeyev2023,kbashi2019,kbashi2018,sergeyev2021,sergeyev2014,sergeyev2014b}). The model describes the evolution of the SOP of the lasing averaged over the pulse width driven by in-cavity CW signal with periodically evolving orthogonal states of polarization:

\begin{equation}
\begin{aligned}
E_x = a \cdot \cos(\Omega t + \varphi_0), \quad\\ 
E_y = a \cdot \sin(\Omega t + \varphi_0) \cdot \exp(i \cdot \Delta\Phi).
\end{aligned}
\end{equation}

Here, $a$ is the amplitude of the injected optical signal, $\Omega$ is the frequency of oscillations, $\varphi_0$ is the initial phase, and $\Delta\Phi$ is the phase difference between the orthogonal SOPs. As follows from Eq.~(4), the injected signal is CW, i.e., the total power is given by $I = I_x + I_y = |E_x|^2 + |E_y|^2 = a^2 = \text{const}$.

To gain an in-depth understanding, we further carried out a linear stability analysis (see Supplementary Materials) to aid the identification of NPR-based oscillations with a frequency of approximately $f_{\text{NPR}} = 4 \times 10^{-3} \cdot f_0$, which closely matches the experimentally observed results (Fig.~\ref{fig:1}c, Fig.~\ref{fig:2}c, and Fig.~\ref{fig:3}c).

\begin{figure}[h]
\centering
\includegraphics[width=0.9\textwidth]{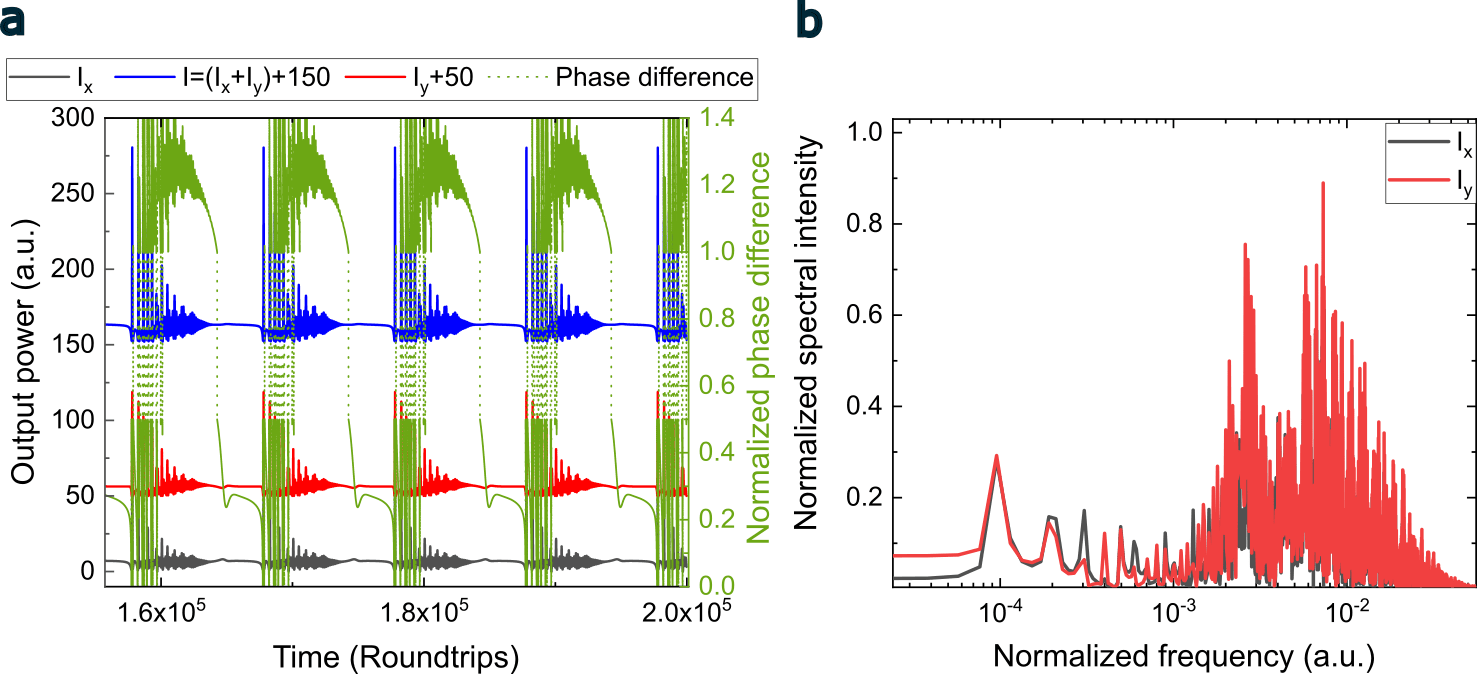}  
\caption{  \textbf{Theoretical results on vector SHE dynamics.}  (a) Polarization dynamics; 
(b) Fast Fourier transform spectrum (see notations in Fig. 2). Parameters: injected signal with the phase difference of $\Delta\Phi = -\pi/4$ and amplitude $a = 0.19$.} 
\label{fig:5}
\end{figure}

By adding a CW signal into the cavity with parameters $a = 0.19$, $\Omega = 2\pi \cdot 0.00005$, $\varphi_0 = \pi/4$, and $\Delta\Psi = -\pi/4$, the QSML dynamics emerge, as shown in Fig.~\ref{fig:5}.

\begin{figure}[h]
\centering
\includegraphics[width=0.9\textwidth]{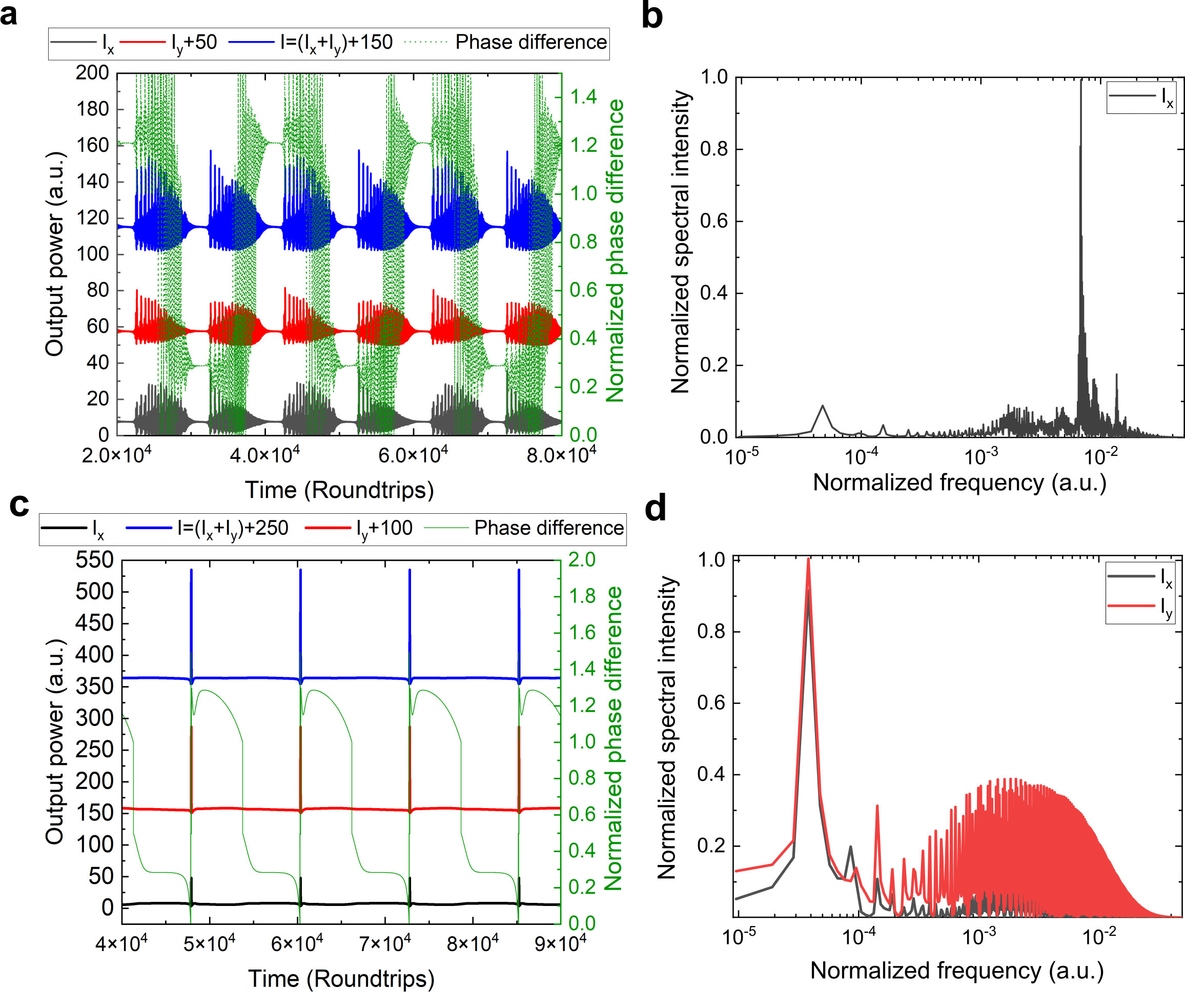}  
\caption{  
\textbf{Theoretical results on vector SHE dynamics.}  a) Polarization dynamics;
b) Fast Fourier transform spectrum (see notations in Fig. 3). Parameters:  injected signal with the phase difference $\Delta\Phi = -\pi/2$ and amplitude $a = 0.185$.
c) Polarization dynamics and d) Fast Fourier transform spectrum for synchronization vs vector harmonic entrainment for the strong coupling of low- and high-frequency oscillations. Notations are the same as for Fig.\ref{fig:1}. Parameters: injected signal with the phase difference $\Delta\Phi = -\pi/4$ and $a = 0.5$.  
}
\label{fig:6}
\end{figure}

As follows from  Fig.\ref{fig:6} c and d, the QSML dynamics of the output powers for polarization modes and total is similar to the experimentally observed (Fig.\ref{fig:2}~c and Fig.\ref{fig:3}~b) in the context of the shape of pulses, the two-scale oscillations in the range of frequencies with $10^{-4} \cdot f_0$ and $10^{-3} \cdot f_0$, and the number of sidebands (Fig.\ref{fig:5}~b), and the phase difference slips of about $\pi$ radians. The phase difference oscillations in Fig.\ref{fig:5}~a  are just slightly different as compared to the experimental results (Fig.\ref{fig:2}~c and Fig.\ref{fig:3}~b).

To specify the proper phase difference between the orthogonal SOPs $\Delta\Phi$, we considered an additional case with $\Delta\Phi = -\pi/2$. The breathing dynamics for this case are shown in Fig.~\ref{fig:6} a, b.

The polarization dynamics occur at low and high frequency oscillations with $f_1 = 5.5 \cdot 10^{-5}$ and $f_2 = 7 \cdot 10^{-3}$. As follows from Fig.~\ref{fig:6}, the breathing dynamics are quite similar to those experimentally and theoretically observed (see Fig.~\ref{fig:3} and Fig.~\ref{fig:5}). The main difference is in the dynamics of the phase difference $\Delta\varphi$ switching. Therefore, the case of $\Delta\Phi = -\pi/4$ presented in Fig.~\ref{fig:6} provides better correspondence to the experimental data shown in Fig.~\ref{fig:2} c, d and Fig.~\ref{fig:3} b, c.

With increased amplitude of the injected signal to $a = 0.5$, the QSML dynamics transform into a spiking regime with frequency $f_1 = 10^{-4}$, as shown in Fig.~~\ref{fig:6} c, d.

\section{Discussion}\
 We revealed experimentally and theoretically that a new vector type of subharmonic entrainment causes double-scale mode-locking dynamics. Unlike the previous study of SHE related to the synchronization of the subharmonic of the fundamental frequency and NPR-driven oscillations \cite{xian2020,wu2023}, it was shown both experimentally and theoretically that vector SHE takes place through overlapping and synchronization of the sidebands of the NPR with low-frequency oscillations having evolving state of polarization. The observed SHE represents a distinct synchronization phenomenon where, as shown in \cite{pikovsky2001,wu2023}, the synchronization behavior depends on both the injected signal amplitude (or coupling coefficient) and the detuning between the injected modulated frequency and the internal oscillation frequency of the system. According to the synchronization theory for subharmonic entrainment in our case, the ratio of frequencies can slightly deviate from an integer, which results in the synchronization in the form of phase difference entrainment, e.g., the phase difference oscillations shown in \cite{pikovsky2001,ding2009,wu2023} and Figs.~\ref{fig:2}, \ref{fig:3}, and \ref{fig:5}. With an increased amplitude of the injected signal (see Fig. \ref{fig:5}), in line with the synchronization theory \cite{pikovsky2001,ding2009,wu2023}, we observe phase- and frequency-locking towards high-power oscillations. To obtain experimentally such dynamics, it is necessary to adjust the power and wavelength of the internal low-frequency oscillations, which is rather cumbersome for discussion here and will be published elsewhere. 

This phenomenon is not only of fundamental interest for laser dynamics but also suggests new avenues for photonic control \cite{thevenin2011}. By tuning the injection parameters—frequency, polarization modulation, and amplitude—one can selectively control envelope timing and polarization state. This could prove useful in ultrafast optical communication, pulse shaping, and metrology, where structured pulse sequences and polarization encoding are advantageous. Crucially, we demonstrate that subharmonic synchronization can occur even when the frequency ratio is not an exact integer. 

In summary, vector subharmonic entrainment provides a new lens through which to view and manipulate laser dynamics. The ability to externally control dual-frequency polarization-encoded pulse trains opens new possibilities for advanced photonic systems and invites further exploration into the interplay of vector dynamics and synchronization. Our findings also connect to broader themes in nonlinear science, including synchronization of high-dimensional and vectorial systems. The analogy between polarization modes in lasers and multi-dimensional coupled oscillators offer a rich framework for exploring complex dynamics, with potential implications for understanding synchronization in biological, chemical, and engineered systems \cite{bialek2012,papo2014,schintler2005,batool2017,okeeffe2017,mancini2010,minati2019,arenas2008,pikovsky2001}.

\section{Methods}\
\textbf{Experimental setup}

The system utilizes a short piece of single-mode highly Erbium-doped optical fiber (EDF) with a peak core absorption of around 110 dB/m at 1530 nm, cutoff wavelength of 890 nm, the numerical aperture of 0.2. The EDF is pumped by a 976-nm pigtailed laser diode (pump laser diode) through a 980/1550 nm wavelength division multiplexer (WDM). The EDF is subsequently spliced to the output port of a dual-stage polarization-sensitive optical isolator (polarization-sensitive ISO) with a center wavelength of 1550 nm and an extinction ratio of 28 dB. The 70/30 optical coupler (OC), featuring a coupling ratio centered at 1550 nm and an insertion loss around 1 $\text{dB}$, was employed to extract $30\%$ of the laser power from the cavity. In addition, the configuration includes two fiber polarization controllers placed before and after the ISO. The first one is the in-line miniature manual in-line fiber polarization controller (FPC). The second one is the electronic in-line polarization controller (EPC) with a three-section external voltage control. The EPC software enables full state-of-polarization control through thermal technology, ensuring stable polarization scanning across the entire Poincaré sphere.  It is important to note that, except for the ISO and EDF, the full cavity of the laser was constructed using standard fiber SMF-28. The total fiber laser cavity length is 12.3 m comprising 0.2 m EDF, 1.5 m PM-1550 fiber, and 10.5 m of SMF-28 fiber. This length also includes pigtail fibers from the EPC, isolator, and coupler. The anomalous group velocity dispersion (GVD) values of the SMF-28 and PM-1550 fibers at $1.55\,\mu\text{m}$ were approximately $- 23\,\text{ps}^2/\text{km}$ and $-22\,\text{ps}^2/\text{km}$, respectively. The normal GVD of the Er-doped fiber at $1550\,\text{nm}$ was estimated to be $15.3\,\text{ps}^2/\text{km}$. As a result, the net dispersion of the cavity could be calculated as $- 0.27\,\text{ps}^2$, suggesting that the laser was operating at an anomalous dispersion regime. For optical signal injection, a Polarization Scrambler (PS) EPS1000 (Novoptel) and a narrow linewidth continuous wave laser (CoBrite DX1, IDPhotonics) are employed.
The output light pulses were characterized and analyzed using various detection and measurement techniques, including a 50 GHz High-Speed Photodetector, a real-time oscilloscope with 6 GHz bandwidth, and an optical spectrum analyzer AQ6317B (Yokogawa).

\textbf{Vector model of NPR mode locked fiber laser with injected signal }

The model includes a distributed form of Jones matrix comprising a combination of Jones matrices for two polarization controllers, (FPC and EPC), with a polariser (ISO), where $\xi_{1(2)}$ is the angle of rotation of the vertical birefringent axis and $\varphi_{1(2)}$ is the phase shift between the wave components in the two orthogonal birefringent axes. The angle between the axis of the polarizer and the $y$ axis is $\theta$.

Considering Eq.~(4), the evolution equations for the complex amplitudes of the lasing field, averaged over the pulse width, and for the population inversion are as follows:

\begin{gather}
\begin{aligned}
\frac{du}{dt_s} &= i \frac{\gamma}{2} \left( |u|^2 u + \frac{2}{3} |v|^2 u + \frac{1}{3} v^2 u^* \right) 
+ (D_{xx} + A_{11}) u + (D_{xy} + A_{12}) v + E_x, \\
\frac{dv}{dt_s} &= i \frac{\gamma}{2} \left( |v|^2 v + \frac{2}{3} |u|^2 v + \frac{1}{3} u^2 v^* \right)
+ (D_{xy} + A_{21}) u + (D_{yy} + A_{22}) v + E_y, \\
D_{xx} &= \left( \frac{a_1 (1 - i \Delta)}{1 + \Delta^2} (f_1 + f_2) - a_2 \right), \quad
D_{xy} = D_{yx} = \left( \frac{a_1 (1 - i \Delta)}{1 + \Delta^2} f_3 \right), \\
D_{yy} &= \left( \frac{a_1 (1 - i \Delta)}{1 + \Delta^2} (f_1 - f_2) - a_2 \right), \\
\frac{df_1}{dt_s} &= \varepsilon (b_1 - a_1 f_1 - a_2 f_2 - a_3 f_3), \\
\frac{df_2}{dt_s} &= \varepsilon \left( b_2 - \frac{a_2}{2} f_1 - a_1 f_2 \right), \\
\frac{df_3}{dt_s} &= - \varepsilon \left( \frac{a_3}{2} f_1 + a_1 f_3 \right), \\
a_1 &= 1 + \frac{I_p}{2} + \frac{\chi (|u|^2 + |v|^2)}{1 + \Delta^2}, \\
a_2 &= \frac{\chi (|u|^2 - |v|^2)}{1 + \Delta^2} + \frac{I_p (1 - \delta^2)}{1 + \delta^2}, \\
a_3 &= \frac{\chi (u \cdot v^* + \text{c.c.})}{1 + \Delta^2}, \\
b_1 &= \frac{(\chi - 1)}{2} I_p - 1, \quad 
b_2 = \frac{I_p}{4} (\chi - 1) \frac{(1 - \delta^2)}{1 + \delta^2}.
\end{aligned}
\label{eq:5}
\end{gather}
Where the transfer matrix for the combination of two polarization controllers, POC1 and POC2 (corresponding to Polarization Controllers FPC and EOPC in Fig. 1, respectively; see supplementary materials for a more detailed scheme Figs.1), and a polarizer (POL, which in this case corresponds to the isolator in Fig. 1) is presented \cite{dengra2012,ding2009}:
\begin{equation}
\begin{aligned}
A &= \ln(T) = 
\begin{bmatrix}
A_{11} & A_{12} \\
A_{21} & A_{22}
\end{bmatrix}, \quad
T = T_{\text{POC1}} \cdot T_{\text{POL}} \cdot T_{\text{POC2}}, \quad i, j = 1, 2. \\[2ex]
T_{\text{POC1(2)}} &= 
\left[
\begin{array}{cc}
\exp\left( \dfrac{i \varphi_{1(2)}}{2} \right) \cos(\xi_{1(2)}) & 
\exp\left( \dfrac{i \varphi_{1(2)}}{2} \right) \sin(\xi_{1(2)}) \\
-\exp\left( -\dfrac{i \varphi_{1(2)}}{2} \right) \sin(\xi_{1(2)}) & 
\exp\left( -\dfrac{i \varphi_{1(2)}}{2} \right) \cos(\xi_{1(2)})
\end{array}
\right], \\[2ex]
T_{\text{POL}} &= 
\left[
\begin{array}{cc}
\sin^2(\theta) & -\dfrac{\sin(2\theta)}{2} \\
\dfrac{\sin(2\theta)}{2} & \cos^2(\theta)
\end{array}
\right].
\end{aligned} 
\end{equation}

Here, time $t_s$ is the slow time normalized to the round-trip time $\tau_r, u = |u|\cdot\exp(i\cdot\varphi_x), v = |v|\cdot\exp(i\cdot\varphi_y), \Delta\varphi = \varphi_y - \varphi_x, |u|^2 = I_x, |v|^2 = I_y$ are normalized to the saturation power $I_{ss}$ and $I_p$ is normalized to the saturation power $I_{ps}$, $\alpha_1$ is the total absorption of Erbium ions at the lasing wavelength; $\alpha_2$ represents the normalized losses; $\delta$ is the ellipticity of the pump wave, $\gamma$ is normalized to the cavity length and the saturation power the Kerr constant, $\varepsilon = \tau_R / \tau_{Er}$ is the ratio of the round-trip time $\tau_R$ to the lifetime of erbium ions at the first excited level $\tau_{Er}$.\\
   $\chi_{(p,s)} = ({\sigma_a^{(s,p)} + \sigma_e^{(s,p)}})/{\sigma_a^{(s,p)}}$, $\sigma_a^{(s,p)}$ and $\sigma_e^{(s,p)}$ are absorption and emission cross-sections at the lasing (s) and pump (p) wavelengths); $\Delta$ is the detuning of the lasing wavelength with respect to the maximum of the gain spectrum (normalized to the gain spectral width); $T_{POC1(2)}$ and $T_{POL}$ are Jones matrices describing SOP transformation by POC1, POC2 and polariser, where $\xi_{1(2)}$ is the angle of rotation of the vertical birefringent axis and $\varphi_{1(2)}$ is the phase shift between the wave components in the two orthogonal birefringent axes, the angle between the axis of the polarizer and the y axis is $\theta$. Matrix $A$ in Eq.4 accounts for the NPR of the lasing field. The functions $f_1, f_2, f_3$ are related to the angular distribution of the excited ions $n(\theta)$ expanded into a Fourier series as follows \cite{sergeyev2023,zeghlache1995}:

\begin{equation}
\begin{aligned}
n(\theta) &= \frac{n_0}{2} \sum_{k=1}^{\infty} n_{1k} \cos(k\theta) + \sum_{k=1}^{\infty} n_{2k} \sin(k\theta), \\[1.5ex]
f_1 &= \left( \chi \frac{n_0}{2} - 1 \right), \quad
f_2 = \chi \frac{n_{12}}{2}, \quad
f_3 = \chi \frac{n_{22}}{2}.
\end{aligned}
\tag{7}
\end{equation}
Equations (5) have been derived by using the approximation that the dipole moments of the absorption and emission transitions for Er-doped silica are located in the plane orthogonal to the direction of light propagation~\cite{sergeyev2023,zeghlache1995}. In contrast to the more general assumption of the 3D distribution of the dipoles’ orientations~\cite{leners1995,sergeyev1999}, an approximation (7) allows getting the finite dimension system presented by Eqs.~(5) where only $n_0$, $n_{12}$ and $n_{21}$ contribute to the dynamics of the lasing field.

We started our study with the case when the polarizer is oriented at angle $\theta = -\pi/4$ and there is no birefringence in the cavity, e.g., $\varphi_{1(2)} = \xi_{1(2)} = 0$ and $A_{11} = A_{12} = A_{21} = A_{22} = -0.6931$. For the case when there is no CW signal in the cavity ($a = 0$), we substitute $u = |u| \exp(i \cdot \varphi_x)$ and $v = |v| \exp(i \cdot \varphi_y)$ into Eq.\ref{eq:5}, and find the steady-state solutions $|u| = |v| = E_{s0}, (E_{s\pi})$, $\Delta\varphi = \varphi_y - \varphi_x = 0, (\pi)$, $f_{1s} \neq 0$, $f_{2s} = 0$, $f_{3s} \neq 0$ as shown in Fig.\ref{fig:4} a, b (details are found in Supplementary Material~\cite{supplementary2025}).

\begin{figure}[h]
\centering
\includegraphics[width=0.9\textwidth]{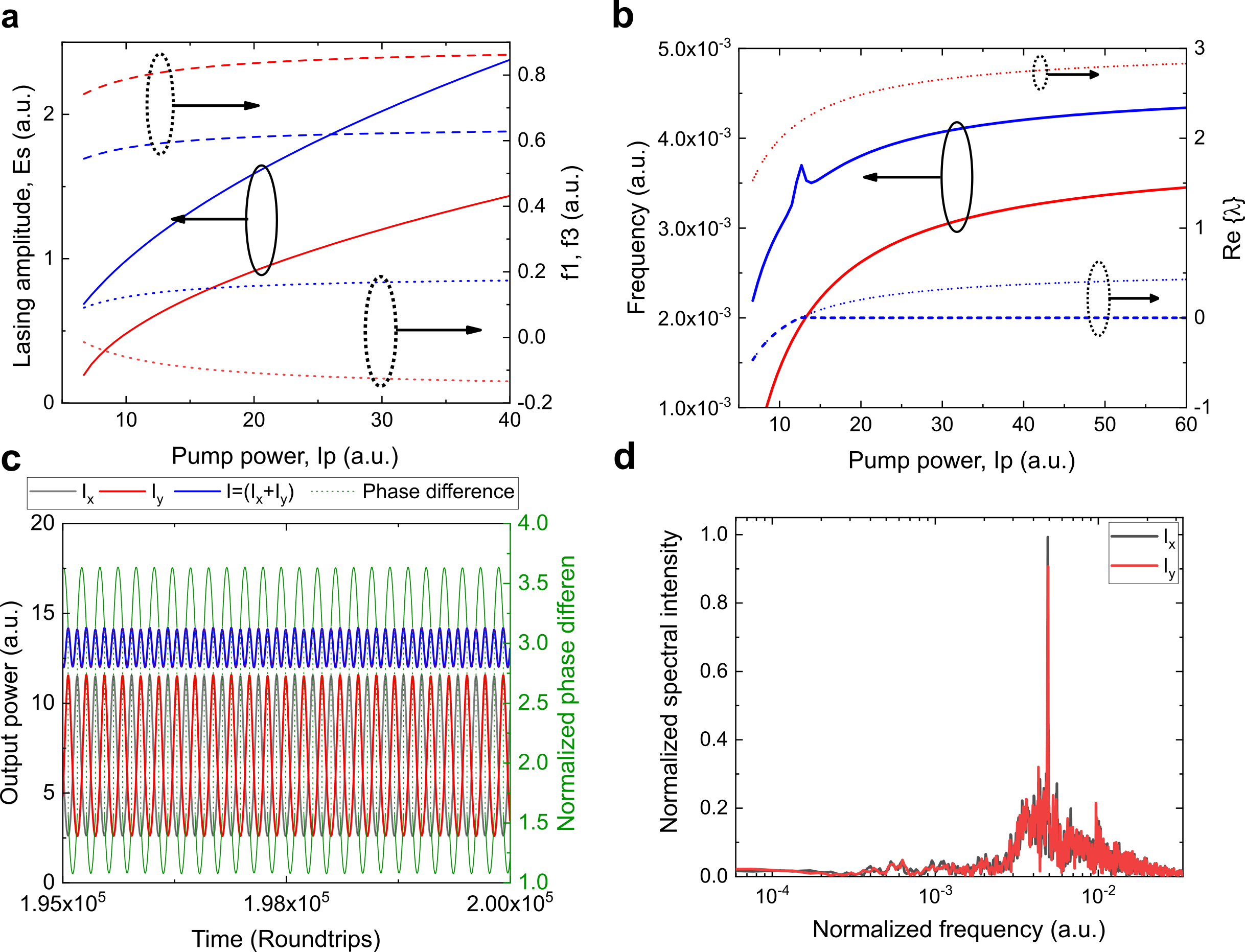}  
\caption{  Steady-state stability analysis of NPR-mode-locked laser without CW signal in the laser cavity.  
(a) Steady-state values of the lasing field as a function of the pump power $I_p$: $\Delta\varphi = 0$ (red colored lines), $\Delta\varphi = \pi$ (blue colored lines), $E_s$ (solid lines), $f_{1s}$ (dashed lines), $f_{3s}$ (dotted lines);  
(b) Imaginary parts of the eigenvalues (frequencies $\Omega_{3,4}$ shown as solid lines) with positive real parts ($R_3$ – dotted lines) and eigenvalue $R_2 > 0$ ($\Delta\varphi = \pi$) shown as dashed line as a function of the pump power $I_p$, $\Delta\varphi = 0$ (red line), $\Delta\varphi = \pi$ (blue line);  
(c) Polarization dynamics obtained by numerical solution of Eqs.~(5);  
(d) Fast Fourier transform spectrum (see notations in Fig.~1).  
Parameters (a–d): $\alpha_1 = 10.131$, $\varepsilon = 0.6 \cdot 10^{-5}$; $\alpha_2 = 2.3$, $\Delta = 0.015$, $\chi = 2.3$, $\gamma = 2 \cdot 10^{-6}$, $\delta = 1$; (c, d): $I_p = 45$,
}
\label{fig:4}
\end{figure}
As follows from Fig. \ref{fig:4} a, the presence of two polarization controllers and polarizer in the cavity result in non-equal lasing field amplitudes for two orthogonal states of polarization with $\Delta\varphi = 0, \pi$.

To specify conditions for the NPR-based oscillations in the case of absence of CW laser radiation in the cavity, we linearize the Eq.\ref{eq:5} in the vicinity of the steady-state solutions mentioned above and find eigenvalues split in two branches I and II as follows (details are found in Supplementary Material).

\begin{equation}
\begin{array}{rl}
(\text{I}) & \lambda_0 = R_0, \quad \lambda_{1,2} = R_1 \pm i\Omega_{1,2}, \\
(\text{II}) & \lambda_3 = R_2, \quad \lambda_{4,5} = R_3 \pm i\Omega_{3,4}, \\
           & \multicolumn{1}{c}{{ R_0} < 0, \quad R_1 < 0,} \\
           & \multicolumn{1}{c}{R_2 < 0\ (\Delta\varphi = 0),} \\
           & \multicolumn{1}{c}{R_2 > 0\ (\Delta\varphi = \pi), \quad R_3 > 0.}
\end{array}
\tag{8}
\end{equation}

The results for $\lambda_{4,5}$ in the case of $\Delta\varphi = 0, (\pi)$ and $\lambda_3$ ($\Delta\varphi = \pi$) are shown in Fig.\ref{fig:4} b. As a result of non-equal amplitudes of the lasing fields for the orthogonal states with $\Delta\varphi = 0, \pi$ (Fig.\ref{fig:4} a), the frequencies of oscillations nearby these states will be non-equal as well (Fig.\ref{fig:4} b). Given the superposition of in-phase ($\Delta\varphi = 0$) and anti-phase oscillations ($\Delta\varphi = \pi$), the numerical solution of Eq.\ref{eq:3}  shows that the total output power will be oscillating as well (Fig.\ref{fig:4} c). 
This modeling framework enables a comprehensive understanding of the vectorial dynamics governing VSHE and confirms the feasibility of entrainment and synchronization  under realistic experimental conditions.

\backmatter

\bmhead{Acknowledgements}

We acknowledge support from Leverhulme Trust grant HARVEST RPG-2023-073, UK EPSRC (EP/W002868/1), Horizon 2020 ETN MEFISTA (861152), National Natural Science Foundation of China (62135007),
National Natural Science Foundation of China (61975107) and 
Natural Science Foundation of Shanghai Municipality (24ZR1422000)
“111” project (D20031). 

\bmhead{Author contributions}
DS and SS planned the project. DS carried out the experiment with helps from SS. DS and SS performed the data acquisition and edited the experimental figures. SS provided the theoretical
support with helps from HK. DS, SS and CM wrote the manuscript with discussions with HK, FW and QH.

\bmhead{Competing interests}

The authors declare no competing interests.

\bmhead{Supplementary information}

 Supplementary material is available for this paper. It includes additional modelling data and a steady-state stability analysis of the NPR mode-locked laser without a CW signal in the laser cavity.
  Correspondence and requests for materials should be addressed to
Dmitrii Stoliarov






\bibliography{sn-bibliography}

\end{document}